\documentclass[11pt, a4paper, twoside]{article}

\usepackage{fancyhdr}
\usepackage[hmarginratio=1:1]{geometry}

\usepackage{amssymb,amsmath,xspace}
\usepackage[breaklinks=true]{hyperref}
\usepackage{graphics}
\usepackage[pdftex]{graphicx}
\usepackage{caption}
\usepackage{subcaption}
\usepackage{enumerate}
\usepackage{booktabs} 
\usepackage{multirow}
\usepackage{float}
\restylefloat{table}
\usepackage{wrapfig,lipsum}

\usepackage{indentfirst}
\usepackage{multirow}
\usepackage{array}
\usepackage[T1]{fontenc}
\usepackage[utf8]{inputenc}
\usepackage{tabularx,ragged2e}
\usepackage[numbers]{natbib}

\newcolumntype{L}[1]{>{\raggedright\let\newline\\\arraybackslash\hspace{0pt}}m{#1}}
\newcolumntype{C}[1]{>{\centering\let\newline\\\arraybackslash\hspace{0pt}}m{#1}}
\newcolumntype{R}[1]{>{\raggedleft\let\newline\\\arraybackslash\hspace{0pt}}m{#1}}

\usepackage{array}

\makeatletter

\newcommand{\thickhline}{%
    \noalign {\ifnum 0=`}\fi \hrule height 2pt
    \futurelet \reserved@a \@xhline
}

\usepackage{abstract}

\title{An Investigation of Methods for Handling Missing Data with Penalized Regression}

\author{Yunjin Choi \vspace{4mm} 
	\\
	\small \it{Department of Statistics} \\
	\small \it{Sequoia Hall, Stanford University, CA, 94305.} \\
	\small \it{{email:}} \textnormal{\texttt{yunjin@stanford.edu}} \vspace{5mm}\\
    Robert Tibshirani \vspace{4mm} \\
	\small \it{Department of Statistics} \\
	\small \it{Sequoia Hall, Stanford University, CA, 94305.} \\
	\small \it{{email:}} \textnormal{\texttt{tibs@stanford.edu}} \vspace{5mm}
	}

\date{}
\begin{document}
\maketitle
\begin{abstract}
\noindent \textbf{Abstract: }We investigate methods for penalized regression in the presence of missing observations.
This paper introduces a method for estimating the parameters which compensates for the missing observations.
We first, derive an unbiased estimator of the objective function with respect to the missing data and then, modify the criterion to ensure convexity. 
Finally, we extend our approach to a family of models that embraces the mean imputation method.
These approaches are compared to the mean imputation method, one of the simplest methods for dealing with missing observations problem, via simulations.
 We also investigate the problem of making predictions when there are missing values in the  test set.

\end{abstract}

\thispagestyle{empty}
\section{Introduction}
	Incomplete data is often found in real world statistical applications.
 As most  statistical methods are developed on an assumption of complete data,
 it is unclear how to apply a statistical method to a data set with missing values. Various approaches have been developed to deal with this problem\cite{book1}. 
In this paper, we focus on the missing observation problem in penalized regression.\\
\indent First, we introduce an approach using a modified minimization criterion of penalized regression\cite{book2}.
Reviewing the elastic net approach,  given a data matrix $X\in \mathbb{R}^{N\times{p}}$ and a response vector $Y \in \mathbb{R}^{N} $, the obective function is as follows:
\begin{eqnarray}
\min_{\beta\in \mathbb{R}^{p}} \left[\frac{1}{2N}||Y - X\beta||^2+ \lambda_1 ||\beta||_1 + \lambda_2||\beta||_{2}^{2} \right].
\label{criterion}
\end{eqnarray}
 The objective function  involves the  data matrix $X$,
so when there are missing values in $X$, it is difficult to construct a  criterion for estimating $\beta$ in the first place. 
For our approach, assuming the observations are missing at random, 
we utilize an unbiased estimator of (\ref{criterion}). Unlike the
objective function, the unbiased estimator  is not necessarily
convex.
 In this case, we modify the unbiased estimator by adding an appropriate
 amount of $\ell_2$ regularization to make it convex. 
In this way, the computation is simple compared to other imputation
methods, especially when there are numerous missing values. Thus,
estimation using this approach, which in this paper we refer to as  {\em  non-negative
  definite covariance} approach, are mainly compared to
the  mean imputation method since it is one of the simplest.
 We compare the  MSE  of each approach via simulated data. \\
\indent Additionally, we extend our approach to combine the
non-negative definite covariance and the mean imputation methods  by
introducing a balancing parameter between these two approaches. 
As the combined method is a generalized method including the non-negative
definite covariance approach and mean imputation methods,
 it can sometimes yield better results than either one.
 The role of the balancing parameter is investigated by simulated examples.\\
\indent Along with the coefficient parameter estimation, we investigate
practical issues in handling missing observations in a test set. Appropriate ways of running cross validation and predicting from an incomplete test set are discussed with examples.

\section{Review of Penalized Regression} 

Penalized regression is a generalized version of ordinary linear regression. By adding a penalty term to an objective function of linear regression, the resulting estimators have useful properties such as variable selection 
and applicability to singular design matrices. 
Let $X\in \mathbb{R}^{N\times{p}}$ be a data matrix and $Y \in \mathbb{R}^{N} $ be a response vector following the model $E(Y|X=x)=\beta_{0}+\beta^{T}x$. The minimization criterion of penalized regression is as follows:
	\begin{eqnarray}
			\min_{(\beta_{0},\beta)\in \mathbb{R}^{p+1}} \left[\frac{1}{2N}\sum_{i=1}^{N}(y_{i}-\beta_{0}-					x_{i}^{T}\beta)^2+
			\lambda P_{\alpha}(\beta)\right]
		\label{eqn:penalty}
	\end{eqnarray}
where
	\begin{eqnarray*}
		P_{\alpha}(\beta) &=& (1-\alpha)\frac{1}{2}||\beta||^{2}_{l_{2}} + \alpha ||\beta||_{l_{1}}\\
		   				  &=& \sum_{j=1}^{p}\left[\frac{1}{2}(1-\alpha)\beta_{j}^2+\alpha|\beta_{j}|\right].
		\label{eqn:penalt2}
	\end{eqnarray*}
By solving the subgradient equations of (\ref{eqn:penalty}) with respect to $\beta_{j}$, we have
	\begin{eqnarray}
		\tilde{\beta_{j}} \leftarrow \frac{S\left( \frac{1}{N}\sum_{i=1}^{N}x_{ij}(y_{i}-\tilde{y_{i}}^{(j)}),				\lambda \alpha \right)}{\frac{1}{N}\sum_{i=1}^{N}x_{ij}^2+\lambda (1-\alpha)}.
		\label{eqn:update_unstandard}
	\end{eqnarray}
 Here, we used $\tilde{y_{i}}^{(j)}:=\tilde{\beta_{0}}+\sum_{l \neq j}\tilde{\beta_{l}} x_{il}$, the fitted value ignoring the role of $j^{th}$ variable and $S(z,\gamma):=sign(z)(|z|-\gamma)_{+}$, a soft-thresholding operator. $\beta$ can be estimated by the cyclic coordinate descent update using the formula (\ref{eqn:update_unstandard}).\\
\indent Here, note that $\tilde{y_{i}}^{(j)}=\hat{y_{i}}-\tilde{\beta_{j}} x_{ij}$ where $\hat{y_{i}}$ is a fitted value using the full model. 
Thus,
	\begin{eqnarray*} 
		\sum_{i=1}^{N}x_{ij}(y_{i}-\tilde{y_{i}}^{(j)}) &=& \sum_{i=1}^{N}x_{ij}(y_{i}-\hat{y_{i}})+\sum_{i=1}^{N}		 x_{ij}^{2} \tilde{\beta_{j}}\\
&=&  	\langle X^{(j)},Y \rangle - \langle X^{(j)},\hat{Y} \rangle + \tilde{\beta_{j}} \langle 							X^{(j)},X^{(j)} 			\rangle \\
&=& 		\langle X^{(j)},Y \rangle - \sum_{|\tilde{\beta_{k}}|>0} \tilde{\beta_{k}} \langle X^{(j)},X^{(k)} 				\rangle +\tilde{\beta_{j}} ||X^{(j)}||_{l_{2}}^{2}.
	\end{eqnarray*}
where $X^{(j)}$ denotes the $j^{th}$ column of the data matrix $X$. Now, rewriting (\ref{eqn:update_unstandard}) in a covariance sense, it becomes
	\begin{eqnarray*}
		\tilde{\beta_{j}} \leftarrow \frac{S(\frac{1}{N}( \langle X^{(j)},Y \rangle -\sum_{|\tilde{\beta_{k}}|>0} 		\tilde{\beta_{k}} \langle X^{(j)},X^{(k)} \rangle + \tilde{\beta_{j}}||X^{(j)}||_{l_{2}}^{2}),\lambda \alpha)}
		{\frac{1}{N}||X^{(j)}||_{l_{2}}^{2} + \lambda (1-\alpha) }.
		\label{eqn:cov_update}
	\end{eqnarray*}

\section{Penalized Regression with Missing Observations}
\subsection{Existing Methods}
There are several existing methods for handling the missing values
problem not only for the penalized regression but for general
statistical analysis. 
Complete case analysis, mean imputation, likelihood-based methods and low rank matrix completion are popular methods to deal the missing values and each method has its motivations and merits\cite{book1} \cite{candes1}.
We also discuss an approach in \citet{wain} which also is motivated from an unbiased estimator of an objective function for estimating parameters like the non-negative definite covariance approach.
\\
\indent Complete case analysis is one of the most basic ways to
confront the missing value problem. This approach ignores all the data
points containing any missing feature and uses only complete data
points as its inputs\cite{book1}. This method is solid in a sense that
it does not use 
any contaminated data,  but also has the drawback of wasting potentially meaningful information. \\
\indent Along with complete case analysis, mean imputation is popular for its simplicity.  It imputes the mean of all available cases of a feature for the missing observations for that feature\cite{book1}. \\
\indent The Likelihood-based approach, like the mean imputation
method, 
imputes the  missing values in some manner. Assuming  some
distribution for the features, in the likelihood-based approach
missing values 
are imputed using the EM algorithm. Multiple items can be imputed simultaneously in a systematic manner and sometimes this can be computationally expensive depending on the model assumptions\cite{book1}. \\
\indent Instead of imputing missing entries, the low rank matrix
completion method approximates a data matrix based on the
singular value decomposition \cite{candes1}. This method is
appropriate when the positions of missing entries are not too
informative and an original matrix is amenable to low rank approximation.\\
\indent \citet{wain} suggest using a unbiased estimator of an objective function to estimate coefficients $\beta$ in regression when data is partially observed or noisy. This approach provides statistical error bounds of estimated $\hat{\beta}$ and also shows polynomial convergence time to global minimum when the gradient descent algorithm is implemented.\\

\subsection{Non-negative Definite Covariance Approach}

Our approach  uses an unbiased estimator of (\ref{eqn:penalty}) for
estimating a true parameter $\beta$,
 where unbiasedness is with respect to a missing pattern of
 observations. Under common assumptions of missing features, such as
 uniform distribution
 and independence within and between features, calculation of the
 unbiased estimator is straightforward. The unbiased estimator,
 however, can be non-convex
 without extra conditions on $\beta$ and thus inconvenient as an
 optimization criterion.
 We avoid this problem by coercing the estimator of covariance matrix
 $\frac{1}{N}X^{t}X$ to be non-negative definite. Using a
 non-negative definite covariance matrix estimator, 
 the objective function becomes convex and thus is  more attractive for optimization. 

\subsubsection{Unbiased Estimator of the Minimization Criterion}

In this paper, we adopt three basic assumptions of the missing pattern in
our data matrix: the existence of missing observations is independent in both within a column and between feature spaces and is uniformly random within each feature space. To be specific, we define $O \in \mathbb{R}^{N \times p}$ to be an indicator matrix of observations where $N$ and $p$ represent a number of data points and a dimension of feature space respectively:
	\begin{eqnarray*}
		O_{ij} &:=& I_{\{ x_{ij} non-missing \}},\\
		O_{ij} &\sim& \mbox{Uniform and i.i.d for fixed }j, \mbox{ and for }i=1,...,N,\\
		O_{ij} &\mbox{and}& O_{ik} \mbox{ are independent for fixed } i \mbox{ and for } j \neq k \in \{ 1,...,p \}.
	\end{eqnarray*} 
Construction of an unbiased estimator of (\ref{eqn:penalty}) is
simple under these assumptions. 
Given a fully-observed standardized data matrix $X$ and a response
vector $Y$ as in the previous section, we define $Z \in \mathbb{R}^{N
 \times p}$ as an observed data matrix, $N_{j}$ as a number of
observed data points in the $j^{th}$ feature and $N_{jk}$ as a number
of observed data points in both $j^{th}$ and $k^{th}$ features.
 We rewrite these as follows:
	\begin{eqnarray*}
		Z_{ij} := O_{ij} \cdot X_{ij} \mbox{, } N_{j} := \sum_{i=1}^{N}O_{ij} \mbox{ and }N_{jk} := \sum_{i=1}^{N}O_{ij} \cdot O_{ik}.
	\end{eqnarray*}
Then the unbiased estimator of (\ref{eqn:penalty}) with respect to the random variable $O$ is as follows:
	\begin{eqnarray}
		\frac{1}{2} \left(  \beta^{t}C_{ZZ} {\beta}-2 C_{YZ}\beta+||Y||_{l_{2}}^{2} \right) + \lambda P_{\alpha}\left( \beta \right)
		\label{Objective_ftn}
	\end{eqnarray}
where $C_{ZZ} \in \mathbb{R}^{p+1 \times p+1} $ and $C_{YZ} \in \mathbb{R}^{1 \times p} $ such that
	\begin{eqnarray}
		\label{covariance}
		\left[ C_{ZZ} \right]_{ij} = \begin{cases}
						      \langle Z^{(i)},Z^{(j)} \rangle /N_{ij} &\text{if }   i\neq j\\
							 ||Z||_{l_{2}}^{2}/N_{j} &\text{if }   i=j
							\end{cases}
		\mbox{ and }
		\left[C_{YZ} \right]_{j} = \langle Y,Z^{(j)} \rangle /N_{j}.
	\end{eqnarray}
%

\subsubsection{Modification for convexity}
Noting that $C_{ZZ}$ is not necessarily non-negative definite,
(\ref{Objective_ftn}) can be non-ideal for optimization 
without constraining the range of $\beta$. We make $C_{ZZ}$ non-negative definite by adding an additional term, converting (\ref{Objective_ftn}) to be tractable by the second order condition of convexity. Specifically, when $C_{ZZ}$ is negative definite, it is replaced by $C_{ZZ} + \gamma I_{p}$ for $\gamma > \Lambda_{min}$ where $\Lambda_{min}$ is the smallest eigen value of $C_{ZZ}$. The modified objective function is:
	\begin{eqnarray*}
		\frac{1}{2} \left(  \beta^{t} (C_{ZZ} + \gamma I_{p}) {\beta}-2 C_{YZ}\beta+||Y||_{l_{2}}^{2} \right) + \lambda P_{\alpha}\left( \beta \right) \mbox{   for   } \gamma > |\Lambda_{min}|I_{\left\{\Lambda_{min}<0 \right\} }.
	\end{eqnarray*}
After reparameterization, it can be rewritten as
	\begin{eqnarray}
		\frac{1}{2} \left( \beta^{t} C_{ZZ}  {\beta}-2 C_{YZ}\beta+||Y||_{l_{2}}^{2} \right) + 
		\lambda_{1} ||\beta||_{l_{1}} + \lambda_{2} ||\beta||_{l_{2}}^{2}
		\label{Objective_ftn_2}
	\end{eqnarray}
where $\lambda_{2} > \frac{1}{2} |\Lambda_{min}I_{ \left\{ \Lambda_{min}<0 \right\} }|$, or equivalently,
	\begin{eqnarray}
		\frac{1}{2} \left(  \beta^{t} C_{ZZ}  {\beta}-2 C_{YZ}\beta+||Y||_{l_{2}}^{2} \right) + 
		\lambda \left( \alpha ||\beta||_{l_{1}} + \frac{1}{2} (1-\alpha) ||\beta||_{l_{2}}^{2} \right)
		\label{Objective_ftn_3}
	\end{eqnarray}
with $\lambda (1-\alpha) > |\Lambda_{min}I_{ \left\{ \Lambda_{min}<0
  \right\} }|$, $\lambda \leftarrow \lambda + | \gamma |$ and $\alpha
\leftarrow \frac{ \lambda \alpha}{\lambda + |\gamma|}$. One remarkable
thing is that this effort to compensate non-convexity
 in (\ref{Objective_ftn}), has resulted in optimization criterion of
 penalized function again as in (\ref{Objective_ftn_2}) or
 (\ref{Objective_ftn_3}).
 A change from the original criterion (\ref{eqn:penalty}) is the range of regularization parameters. Now, $\beta$ can be estimated by minimizing (\ref{Objective_ftn_3}) using cyclic coordinate descent as in section 2 \cite{regularization, pathwise}:
	\begin{eqnarray}
		\tilde{\beta_{j}} \leftarrow \frac{S( \frac{1}{N_{j}}\langle Z^{(j)},Y \rangle - \sum_{|					\tilde{\beta_{k}}|>0}\frac{\tilde{\beta_{k}}}{N_{jk}} \langle Z^{(j)},Z^{(k)} \rangle, \lambda \alpha)}
                            { \frac{1}{N_{j}} ||Z^{(j)}||_{l_{2}}^{2} + \lambda ( 1-\alpha)}.
	\end{eqnarray}
Note that the meaningful upper bound for $\lambda$ would be $\lambda < \frac{1}{\alpha} \max_{j \in \langle \{1,2,...,p \rangle \}} |\frac{\langle Z^{(j)},Y \rangle}{N_{j}}|$ since beyond this threshold, the estimated $\hat{\beta}$ is estimated to be 0. Combining this with the bound from (\ref{Objective_ftn_3}), the valid range of $\lambda$ and $\alpha$ are as follow:
	\begin{eqnarray*}
		\label{range}
		\lambda \alpha &\in & \left[0, \max_{j \in \{1,...,p \} } | \left[ C_{YZ} \right]_{j} |\right]\\
                   \text{and   } 
		\alpha & \in & \left[ 0, \frac{\max | \left[ C_{YZ} \right]_{j} | }{ | \Lambda_{min}I_{\Lambda_{min} <0} |+ \max | \left[ C_{YZ} \right]_{j} | } \right].
	\end{eqnarray*}

\subsubsection{Test Set Prediction and Cross Validation}
When there are missing observations in a test set, 
 it is unclear how to make a prediction on the set.
 For the same reason,, applying cross validation is problematic. Here
 we impute the incomplete test obsverations using conditional
 expectations. After imputing the incomplete test data, we can apply estimated $\beta$ directly. To be specific, when observations of features $j = j^{1},...,j^{k}$ for $i'^{th}$ data point in test are missing, we used
	\begin{eqnarray}
		\label{condExp}
		\left( \hat{X}_{ij^{1}}^{test},...,\hat{X}_{ij^{k}}^{test} \right)
		= E \left[ (X_{ij^{1}},...,X_{ij^{k}}) \right | 
  		\left\{ X_{ij} | j \neq j^{1},..., j^{k} \right\} ] \mbox{ where }
  		X_{i \cdot} \sim N_{p}(\mu, \Sigma ).
	\end{eqnarray}
We use the training mean for $\mu$ and $C_{zz} + \left(
  |\Lambda_{min}| + \lambda(1-\alpha)\right) I_{p} $ for $\Sigma$
where $\lambda$ and $\alpha$ are regularization parameters in
(\ref{Objective_ftn_3}). 
An extra term $\lambda(1-\alpha)I_p$ is added to $C_{ZZ}$ which is
element-wise unbiased estimator of the true $\Sigma$ to avoid
singularity, since the conditional expectation of multivariate normal
distribution involves an inverse of submatrix of $\Sigma$. For the case
when $\alpha = 1$ and the extra term vanishes, a pseudo inverse is used if a submatrix of interest is singular.

\subsubsection{Comparison of Non-negative Definite Covariance Approach and Mean Imputation}
	\label{simSet}

\begin{figure}[tp]
	\vspace{-0.7cm}
	\begin{tabular}{c}
		\includegraphics[height=2.1in, width=\textwidth]{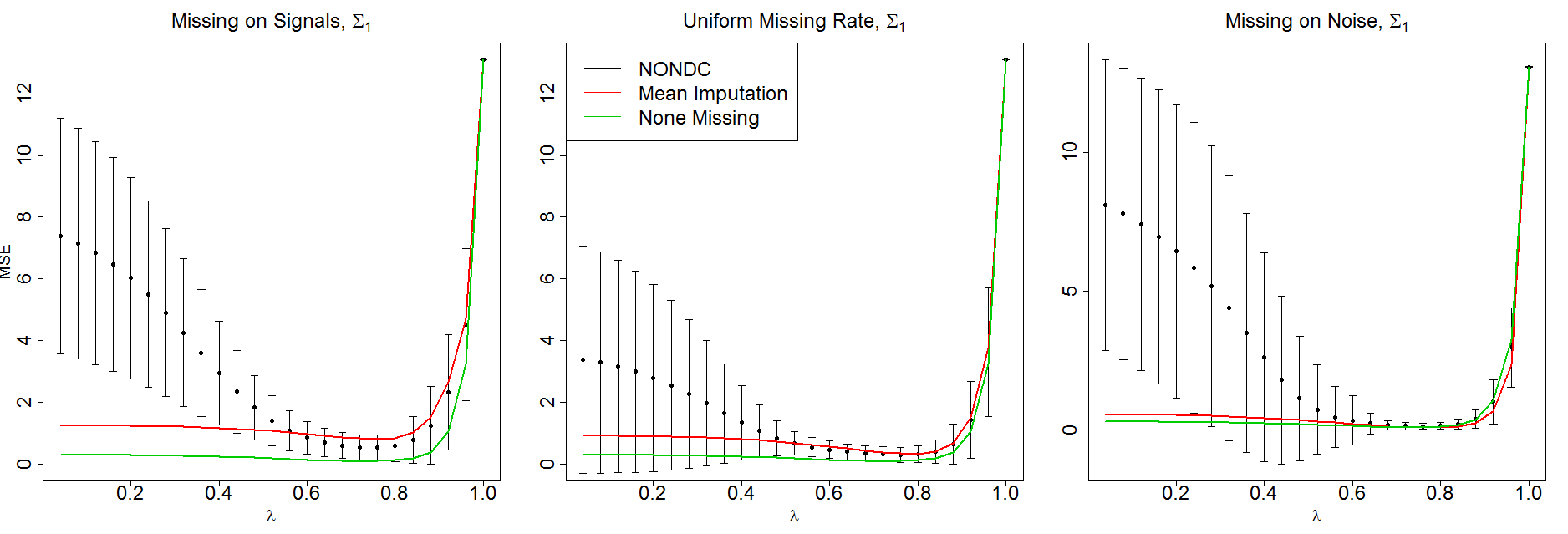} \\
		\includegraphics[height=2.1in, width=\textwidth]{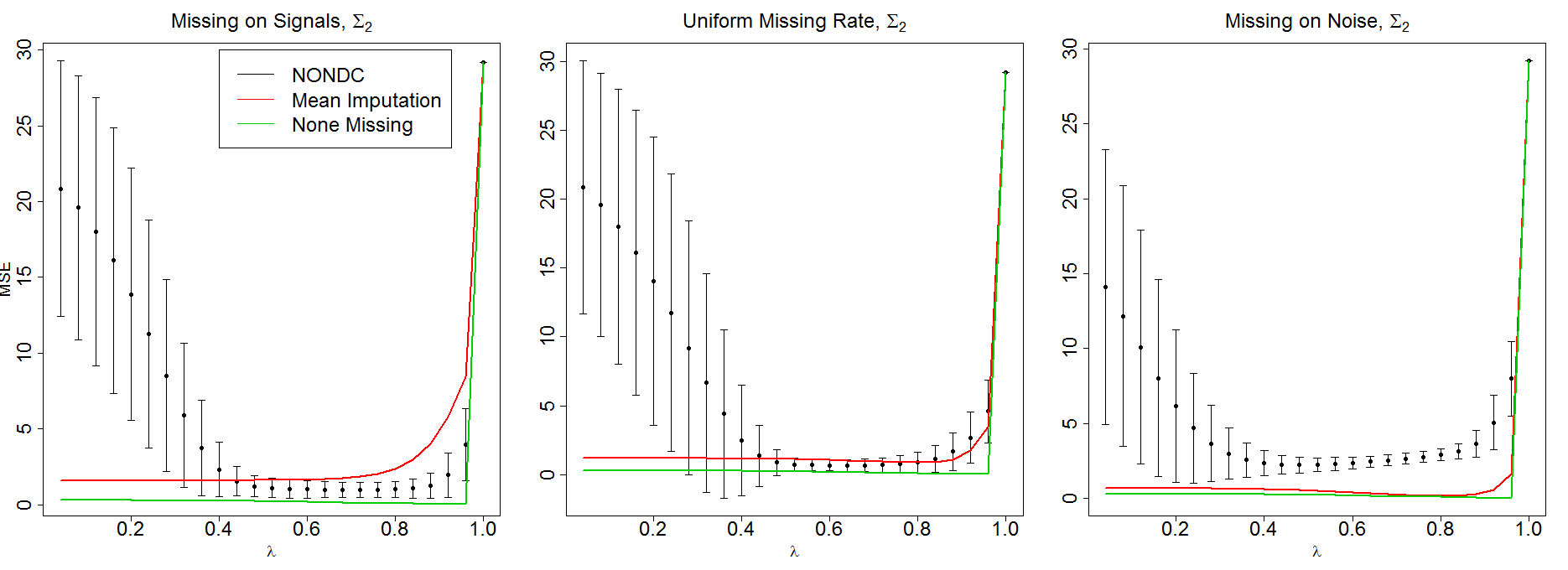} \\
		\includegraphics[height=2.1in, width=\textwidth]{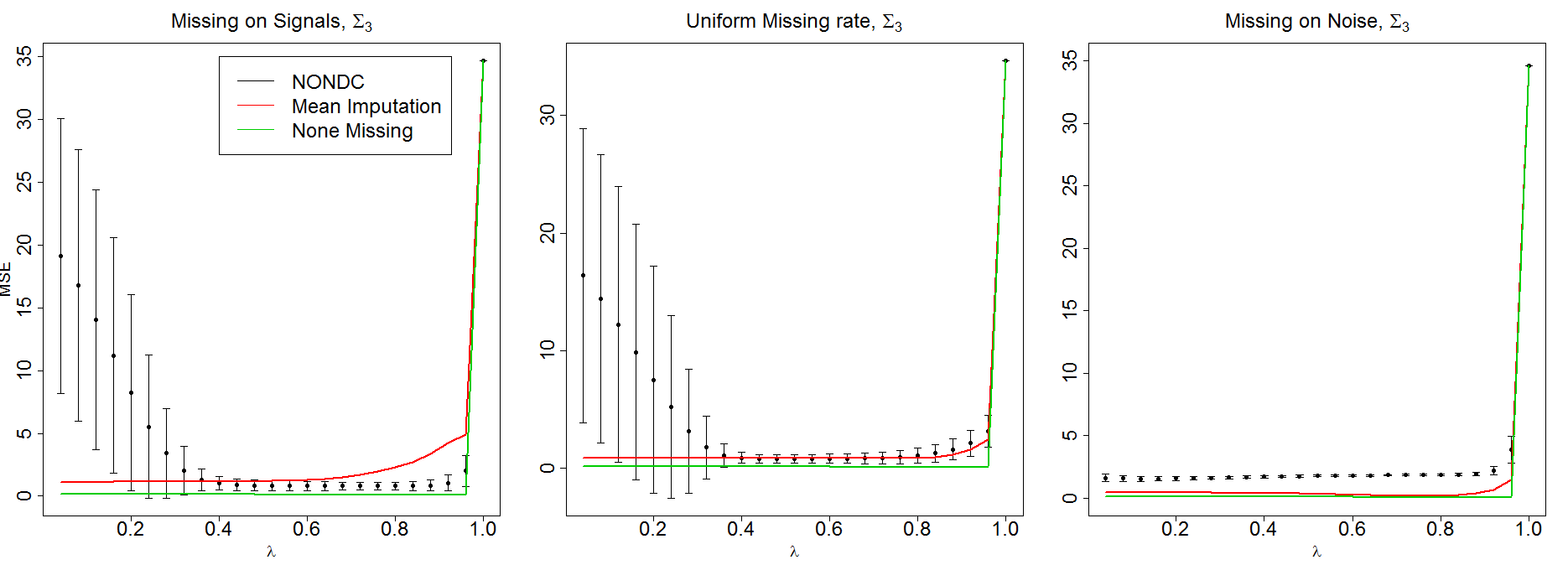} \\
		\includegraphics[height=2.1in, width=\textwidth]{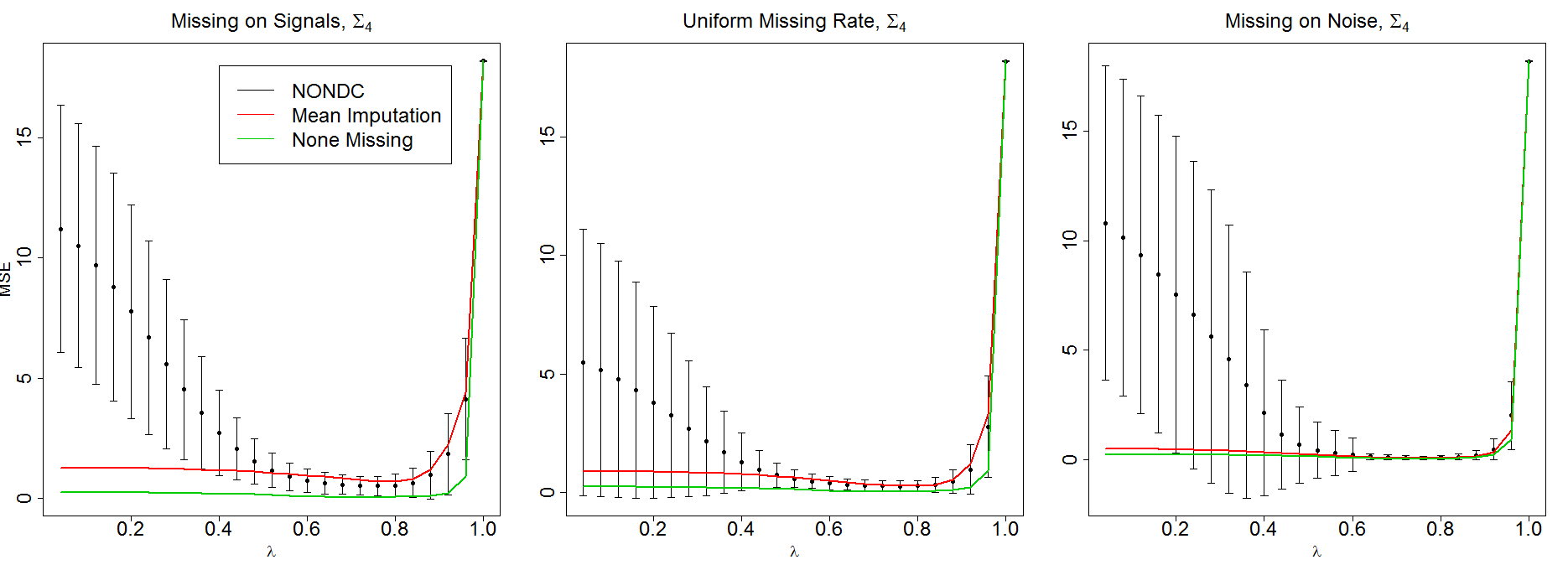} \\
	\end{tabular}
	\caption{\small \it{Plots of estimated MSE of the 12 scenarios over 300 trials. Error bars contain the mean value and $\pm 1$ se of the mean. For each plot, $\alpha$ is fixed at the point which bears the global minimum. The cases in which missing rate is high on signals(left column), the cases in which missing rate is uniform for all covariates(middle column) and the cases in which missing rate is high on dummy variables(right column) are plotted in the figure. The data matrix $X$ in the $i^{th}$ row is generated under multivariate normal distribution with covariance matrix proportional to $\Sigma_{i}$ as in (\ref{structure}).}\label{fig:figure1}}
\end{figure}

In this section, we discuss the performance of the non-negative
definite covariance approach in comparison to mean imputation via
simulated data under various settings. In every instance, the data is generated under a linear model:
	\begin{eqnarray*}
		Y = X \beta + \epsilon \mbox{, } \epsilon \sim N(0, \sigma^2 I_{p})
	\end{eqnarray*}
with fixed $N=50$ and $p=15$ where $X$ is generated under multivariate normal distribution:
	\begin{eqnarray*}
		X \sim N(\mu, \Sigma).
	\end{eqnarray*}
The coefficient $\beta$ is fixed to be $\beta = \left(0, 2, 2,
  2,0,...,0 \right)$ and $\sigma$ is set to have signal-to-ratio of 4.
 The covariates corresponding to non-zero and zero entries in $\beta$
 are considered to be true signals and dummies respectively.
 We investigated 12 scenarios which are combinations of three types of missing pattern and 4 types of $\Sigma$ of a data matrix $X$. \\
\indent For the 3 missing patterns, a case when missing observations
are concentrated on the signals, a case when missing rate is uniform
over all covariates and
 a case when missing observations are concentrated on dummy variables are investigated:
	\begin{equation*} 
		O_{ij} \sim \left\{
		\begin{array}{l l}
		Bernoulli(\gamma) & \mbox{if missing rate is uniform} \\
		Bernoulli(2\gamma \frac{2p-2j+1}{2p}) & \mbox{if missing rate is high on signals} \\
		Bernoulli(2\gamma\frac{2j-1}{2p}) & \mbox{if missing rate is high on dummy variables}. \\
		\end{array}
		\right.
	\end{equation*}
Here $\gamma$ denotes for the average missing rate in each case and we
used $\gamma = 0.25$ for all cases. For $\Sigma$, we tried the four following cases:
	\begin{eqnarray}
		&& \Sigma_{i} = \sigma^2
			\begin{bmatrix}
				1 		 & \rho_{i} & \ldots  & \rho_{i} \\
				\rho_{i} & 1		    & \ldots  & \rho_{i} \\
				\vdots   & \vdots   & \vdots  & \vdots   \\
				\rho_{i} & \ldots   &\rho_{i} & 1 \\
		\end{bmatrix}
				\mbox{ with } \rho_{1} = 0 \mbox{, } \rho_{2} = 0.5 \mbox{, and } \rho_{3} = 0.75 \\	\mbox{and } && \left[ \Sigma_{4} \right]_{ij} = \rho^{|i-j|} \mbox{ with } \rho = 0.5.
	\label{structure}
	\end{eqnarray}  
To investigate the efficacy of the methods, the MSE of $X\hat{\beta}$
was used.
 The MSE [$E[(X\hat{\beta} - X\beta)^2]$] is estimated over 300
 repetitions where the expectation is over an observation $O$ and
 noise $\epsilon$. Figure \ref{fig:figure1} and table \ref{tab:table1}
show that NONDC works better than mean imputation method when missing
rate 
is high on signals while mean imputation method surpasses NONDC when
missing rate is high on dummy variables.
 The non-negative definite covariance technique amplifies values in covariance matrix corresponding to high missing rate by scaling these elements by larger values of $\frac{1}{N_{jk}}$. As a result, when missingness is concentrated on signals, the role of significant covariates is emphasized leading to a good estimation.\\

\begin{table}
\begin{minipage}{\linewidth}
	\centering
	Global Minimum MSE
	
	\bigskip
		\begin{tabular}{|c|c| C{2.5cm} | C{2.5cm} | C{2.5cm}| }
		\hline 
	$\Sigma$	in									  &	\multirow{2}{*}{Approach}			   &	\multicolumn{3}{c|}{Missing rate} \\ 
	\cline{3-5}	
	$X \sim N(\mu, \sigma^2 \Sigma))$ &  					
		& High on signals & 
	   		Uniform 
		& High on dummy variables\\
		\thickhline

		\multirow{2}{*}{$\Sigma_{1}$} & NONDC & 0.53 (0.42) & 0.29 (0.24) & 0.12 (0.12) \\
													& MI			& 0.80 (0.49) & 0.33 (0.27) & 0.08 (0.07) \\
		\hline
		\multirow{2}{*}{$\Sigma_{2}$} & NONDC & 0.97 (0.49) & 0.67 (0.43) & 2.22 (0.53) \\
													& MI			& 1.59 (0.33) & 0.87 (0.44) & 0.16 (0.17) \\
		\hline
		\multirow{2}{*}{$\Sigma_{3}$} & NONDC & 0.80 (0.32) & 0.79 (0.37) & 1.56 (0.19) \\
													& MI			& 1.12 (0.22) & 0.83 (0.26) & 0.19 (0.16) \\
		\hline
    	\multirow{2}{*}{$\Sigma_{4}$} & NONDC & 0.51 (0.40) & 0.27 (0.22) & 0.09 (0.08) \\
													& MI			& 0.70 (0.52) & 0.28 (0.24) & 0.08 (0.06) \\
		\hline
	\end{tabular} 
	\par

\end{minipage}
	\caption{
	Global minimum MSE over regularization parameters $\alpha$ and $\lambda$ of 12 different scenarios. Each cell represents one scenario with two different methods(the non-negative definite covariance and the mean imputation). Non-negative definite covariance approach and mean imputation method are refered to as NONDC and MI respectively. Global minimum MSE for each case is estimated over 300 trials. The values in parenthesis are corresponding 1 se. 	\label{tab:table1}} 
\end{table}

\subsection{A Combined Approach}

	The simulated examples show that neither the non-negative definite covariance approach nor the mean imputation method dominates. 
Thus, an approach embracing both methods might be beneficial. Here we combine these two methods by introducing a new parameter $\eta$ which can be interpreted as a balancing parameter of the two competing methods.
	
\subsubsection{Implementation}

The basis of the non-negative definite covariance approach is to replace the covariance matrix $X^{t}X$ and $X^{t}Y$ by $C_{XX}$ and $C_{XY}$ defined in (\ref{covariance}). The combined approach amends (\ref{covariance}) so that it can embrace the mean method:
\begin{eqnarray*}
\label{newCovariance}
\left[ C_{ZZ}^{\eta} \right]_{ij} = \begin{cases}
						     \left( \frac{ 1-\eta }{N_{ij} }+ \frac{\eta}{ N} \right) \langle Z^{(i)},Z^{(j)} \rangle &\text{if }   i\neq j\\
							 \left( \frac{1 - \eta}{N_{j}}+\frac{\eta}{N} \right) ||Z||_{l_{2}}^{2} &\text{if }   i=j
							\end{cases}
\mbox{ and }
\left[C_{YZ}^{\eta} \right]_{j} = \left( \frac{ 1 - \eta }{N_{j}}+\frac{\eta}{ N} \right) \langle Y,Z^{(j)} \rangle
\mbox{ for } \eta \in [0,1].
\end{eqnarray*}
This is identical to the non-negative definite covariance approach
when $\eta =0$ while it is equivalent to mean imputation for $\eta =
1$.
 For $\eta$ between 0 and 1, this approach inherits advantages of both
 methods.
 As the only changes in the combined method from the non-negative definite covariance method are $C_{ZZ}$ and $C_{ZY}$, we can estimate $\hat{\beta}$ in the same manner as in the non-negative definite covariance approach just by plugging $C_{ZZ}^{\eta}$ and $C_{ZY}^{\eta}$ into the corresponding places in (\ref{Objective_ftn}). Thus the objective function in this combined approach is
\begin{eqnarray*}
\frac{1}{2} \left(  \beta^{t} C_{ZZ}^{\eta}  {\beta}-2 C_{YZ}^{\eta}\beta+||Y||_{l_{2}}^{2} \right) + 
\lambda \left( \alpha ||\beta||_{l_{1}} + \frac{1}{2} (1-\alpha) ||\beta||_{l_{2}}^{2} \right).
\end{eqnarray*}
	Again $\hat{\beta}$ can be estimated by cyclic coordinate descent as follows:
\begin{eqnarray*}
\tilde{\beta_{j}} \leftarrow \frac{S( [C_{ZY}^{\eta}]_{j} - \sum_{|\tilde{\beta_{k}}|>0}\tilde{\beta_{k}}
[C_{ZZ}^{\eta}]_{jk}, \lambda \alpha)}
                            { [C_{ZZ}^{\eta}]_{jj} + \lambda ( 1-\alpha)}.
\end{eqnarray*}
with the range of $\alpha$ and $\lambda$ being
\begin{eqnarray*}
\lambda \alpha \in \left[0, \max_{j \in 1,...,p} | \left[ C_{YZ}^{\eta} \right]_{j} |\right]
                   \text{and   } 
\alpha \in \left[ 0, \frac{\max | \left[ C_{YZ}^{\eta} \right]_{j} | }{ | \Lambda_{min}I_{\Lambda_{min} <0} |+ \max | \left[ C_{YZ}^{\eta} \right]_{j} | } \right]
\end{eqnarray*}
where $\Lambda_{min}$ is the smallest eigen value of $C_{ZZ}^{\eta}$.\\

\indent In the combined method, predicting the values
in an incomplete test set can be conducted in the same manner as in the non-negative definite covariance approach. Like the non-negative definite covariance approach, we use conditional expectation on assuming multivariate normal distribution on a feature space. In the combined method, we estimate $\Sigma$ by $C_{ZZ}^{\eta} + \left( |\Lambda_{min}| + \lambda(1-\alpha)\right) I_{p}$ in which $C_{ZZ}$ in (\ref{condExp}) is replaced by $C_{ZZ}^{\eta}$. Again, when $\alpha=1$ and $\Sigma$ becomes singular, pseudo inverse is used for conditional expectation.

\subsubsection{Simulation Results}
\begin{figure}[p 13]
	\centering
	\includegraphics[scale = 0.45]{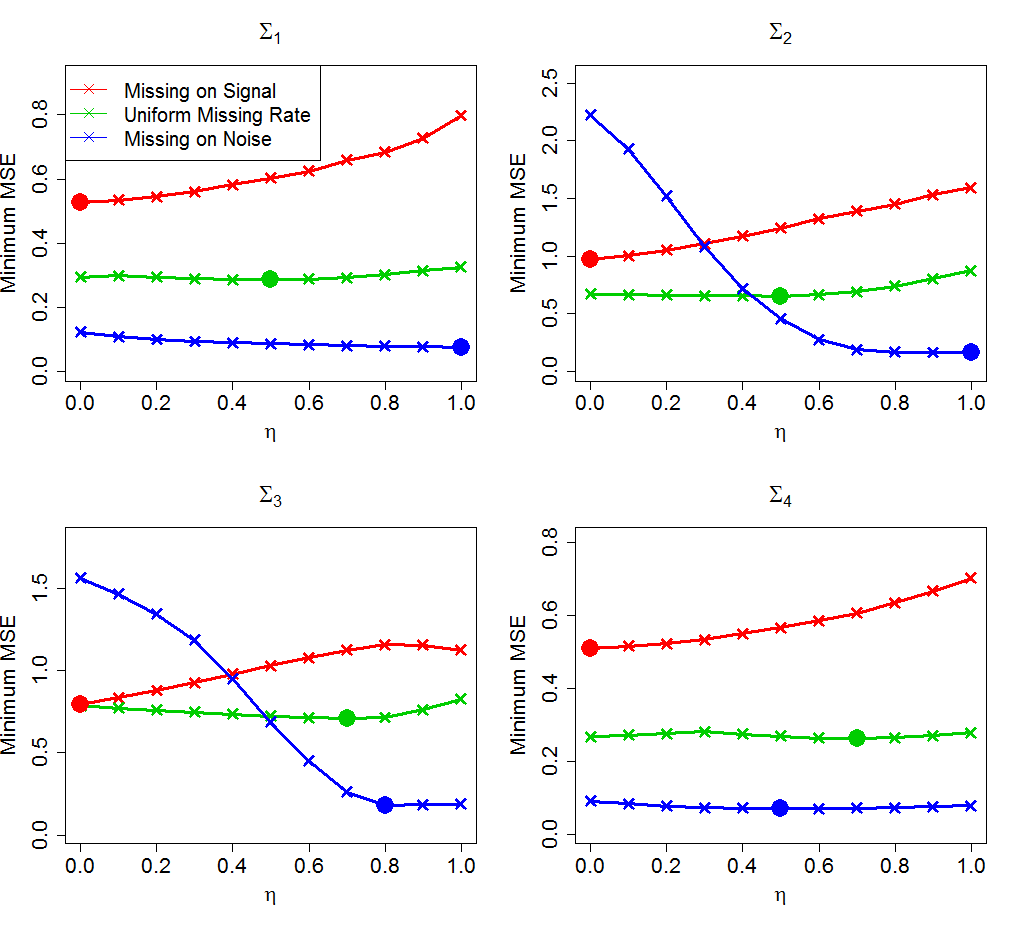}
	\label{incorporate}
	\caption{ \small \it{Plots of estimated minimum MSEs over 300 trials of combination of 4 different data structure and 3 different missing patterns. Each line represents the minimum MSEs of one scenario. The minimum MSE at given $\eta$ is achieved over $\alpha$ and $\lambda$.
	Each plot represents one of the data structures noted in (\ref{structure}). Cases when the covariance matrix of the matrix $X$ is $\Sigma_1$(top left), $\Sigma_2$(top right), $\Sigma_3$(bottom left) or $\Sigma_4$(bottom right) are shown. Lines in each plot represent 3 different missing patterns. The case in which missing rate is high on signals(red line), the case in which missing rate is uniform over features(green line) and the case in which missing rate is high on dummy variables(blue line) are plotted. A solid dot in each line represents the global minimum of the corresponding case.} \label{fig:figure2}}

\end{figure}
 \begin{table}
\begin{minipage}{\linewidth}
	\centering
	Global Minimum MSE
	\bigskip
	\label{tab:orMSE}
		\begin{tabular}{|c|c| C{2.5cm} | C{2.5cm} | C{2.5cm}| }
		\hline 
	$\Sigma$	in									  &	\multirow{2}{*}{Approach}			   &	\multicolumn{3}{c|}{Missing rate} \\ 
	\cline{3-5}	
	$X \sim N(\mu, \sigma^2 \Sigma))$ &  					
		& High on signals & 
	   		Uniform 
		& High on dummy variables\\
		
		\thickhline

		\multirow{3}{*}{$\Sigma_{1}$} 
		
		& Comb  & 0.53 (0.42) & \bf{0.29} (0.25) & 0.08 (0.07) \\		
		& NONDC & \bf{0.53} (0.42) & 0.29 (0.24) & 0.12 (0.12) \\
									  & MI			& 0.80 (0.49) & 0.33 (0.27) & \bf{0.08} (0.07) \\
		\hline
		\multirow{3}{*}{$\Sigma_{2}$} & Comb & 0.97 (0.49) & \bf{0.65 (0.40)} & 0.16 (0.17) \\
									  & NONDC & \bf{0.97 (0.49)} & 0.67 (0.43) & 2.22 (0.53) \\
									  & MI			& 1.59 (0.33) & 0.87 (0.44) & \bf{0.16} (0.17) \\
		\hline
		\multirow{3}{*}{$\Sigma_{3}$} & Comb & 0.80 (0.32) & \bf{0.71} (0.33) & \bf{0.18} (0.15) \\
									  & NONDC &\bf{0.80} (0.32) & 0.79 (0.37) & 1.56 (0.19) \\
									  & MI			& 1.12 (0.22) & 0.83 (0.26) & 0.19 (0.16) \\
		\hline
    	\multirow{3}{*}{$\Sigma_{4}$} & Comb & 0.51 (0.40) & \bf{0.26} (0.23) & \bf{0.07} (0.06)
 \\
    								  & NONDC & \bf{0.51} (0.40) & 0.27 (0.22) & 0.09 (0.08) \\
									  & MI			& 0.70 (0.52) & 0.28 (0.24) & 0.08 (0.06) \\
		\hline
	\end{tabular} 
	\par
	\caption{ \small \it{
	Global minimum MSEs over regularization parameters $\alpha$ and $\lambda$ of 12 different scenarios. Each cell represents one scenario with 3 different methods(the combined approach, the non-negative definite approach and the mean imputation). The combined method, the non-negative definite covariance approach and the mean imputation method are referred to as Comb, NONDC and MI respectively. The Global minimum MSE for each case is estimated over 300 trials. The values in parentheses are corresponding 1 se. The smallest value among three different approaches in a cell is represented in bold letter.} \label{tab:table2}} 
\end{minipage}
\end{table}
\begin{table}[h]
	\vspace{-0.8cm}
\begin{center}
	\centering Test Error
	
		\bigskip
		\begin{tabular}{|c|c| C{2.5cm} | C{2.5cm} | C{2.5cm}| }
		\hline 
		\multirow{2}{*}{Covariance of X}									  &	\multirow{2}{*}{$\Sigma$ used for}			   &	\multicolumn{3}{c|}{Missing rate} \\ 
	\cline{3-5}	
	    &  	test set imputation			
		& High on signals & 
	   		Uniform 
		& High on dummy variables\\
		
		\thickhline

		\multirow{3}{*}{$\Sigma_{1}$} 
		
		& $\Sigma_{est}$  & 6.55 (1.16) & 4.32 (0.91) & 1.40 (0.36) \\		
		& $I$ & \bf{6.23} (1.05) & \bf{3.58} (0.74) & \bf{1.24} (0.33) \\
		& $\Sigma_{true}(=\Sigma_1)$			& 6.23 (1.05) & 3.58 (0.74) & 1.24 (0.33) \\
		\hline
		\multirow{3}{*}{$\Sigma_{2}$} 
		&  $\Sigma_{est}$ & \bf{4.74} (0.80) & 4.39 (1.62) & \bf{1.16} (0.36) \\
		& $I$ & 7.96 (1.34) & \bf{4.11} (0.95) & 1.55 (0.49)
 \\
		& $\Sigma_{true}(=\Sigma_2)$			& 4.40 (0.79) & 2.64 (0.79) & 1.09 (0.34) \\
		\hline 
		\multirow{3}{*}{$\Sigma_{3}$} & $\Sigma_{est}$ & \bf{2.73} (0.45) & \bf{2.67} (0.81) & \bf{0.70} (0.23) \\

									  & $I$ & 6.18 (1.17) & 3.75 (1.05) & 1.17 (0.38) \\
									  & $\Sigma_{true}(=\Sigma_3)$			& 2.54 (0.45) & 1.86 (0.61) & 0.55 (0.17) \\
		\hline
	\end{tabular} 
	\par
	\caption{
	\small \it{Test errors evaluated at optimal $({\alpha}, {\lambda}, {\eta})$ of 9 different scenarios. Each cell represents one scenario with 3 different $\Sigma$ used for an incomplete test set imputation. $\Sigma_{est}$ represents $C_{ZZ}^{{\eta}} + (|\Lambda_{min}| + {\lambda}(1-{\alpha}))I$ which is the suggested method in this paper while $\Sigma_{true}$ denotes the true covariance matrix of the data matrix $X$. $\Sigma_{true}$ and $I$ are presented for reference. The optimal $({\alpha}, {\lambda}, {\eta})$ in each setting is estimated to be the point which yields the global minimum MSE which is estimated over 300 repetitions. Test error values are estimated over 50 trials. The values in parenthesis are corresponding 1 se. The smaller value between $\Sigma_{est}$ and $I$ is represented in bold letter.} \label{tab:table3}} 
\end{center}
\end{table}
\begin{table}[h]
\begin{center}
	\centering Ratio of Global Minimum MSE and MSE at ($\alpha$,$\lambda$,$\eta$) Chosen by Cross Validation
	
	\bigskip
		\begin{tabular}{|c|c| C{2.5cm} | C{2.5cm} | C{2.5cm}| }
		\hline 
		\multirow{2}{*}{Covariance of X}									  &	\multirow{2}{*}{$\Sigma$ used for}			   &	\multicolumn{3}{c|}{Missing rate} \\ 
	\cline{3-5}	
	    &  	test set imputation			
		& High on signals & 
	   		Uniform 
		& High on dummy variables\\
		
		\thickhline

		\multirow{3}{*}{$\Sigma_{1}$} 

		& $\Sigma_{est}$  & \bf{1.55} (0.32) & 1.62 (0.55) & 2.20 (1.37) \\		
		& $I$ & 1.60 (0.29) & \bf{1.49} (0.61) & \bf{1.92} (0.98) \\
		& $\Sigma_{true}(=\Sigma_1)$ & 1.60 (0.29) & 1.49 (0.61) & 1.92 (0.98) \\
		\hline
		\multirow{3}{*}{$\Sigma_{2}$} 
		&  $\Sigma_{est}$ & \bf{1.50} (0.35) & \bf{1.64} (0.47) & \bf{1.49} (0.33) \\
		& $I$ & 2.09 (0.32) & 1.91 (0.68) & 2.01 (0.82) \\
		& $\Sigma_{true}(=\Sigma_2)$			& 1.43 (0.27) & 1.21 (0.23) & 1.49 (0.42) \\
		\hline
		\multirow{3}{*}{$\Sigma_{3}$} & $\Sigma_{est}$ & \bf{1.34} (0.24) & \bf{1.44} (0.24) & \bf{1.18} (0.14) \\
									  & $I$ & 1.69 (0.15) & 1.59 (0.29) & 1.73 (0.60) \\
									  & $\Sigma_{true}(=\Sigma_3)$ & 1.33 (0.25) & 1.19(0.30) & 1.20 (0.17)
 \\
		\hline
	\end{tabular} 
	\par
	\caption{
	\small \it{Ratio of global minimum MSE and MSE at ($\hat{\alpha}$,$\hat{\lambda}$,$\hat{\eta}$) chosen by cross validation of 9 different scenarios. Each cell represents one scenario with 3 different $\Sigma$ used for an incomplete test set imputation. $\Sigma_{est}$ represents $C_{ZZ}^{\hat{\eta}} + (|\Lambda_{min}| + \hat{\lambda}(1-\hat{\alpha}))I$ which is the suggested method in this paper while $\Sigma_{true}$ denotes the true covariance matrix of the data matrix $X$. $\Sigma_{true}$ and $I$ are presented for reference. The ratio values are estimated over 50 trials. The values in parenthesis are corresponding 1 se. The smaller value between $\Sigma_{est}$ and $I$ is represented in bold letter.} \label{tab:table4}}
\end{center}
\end{table}
In this section, we will first compare the performance of the mean imputation method, the non-negative definite covariance approach and the combined method. Second, we will discuss the ability of the combined method in choosing proper regularization parameters using cross validation. Finally, we will evaluate test error values of the combined method. 
The simulation settings in this section are the same as in section~\ref{simSet}. 
For an incomplete test set imputation 
for evaluating both cross validation and test error, we used 3 different types of $\Sigma$
in (\ref{structure}): $\Sigma = C_{ZZ}+\left(|\Lambda_{min}|
  + \lambda(1-\alpha)\right)I$, $\Sigma = I$ and $\Sigma =
\Sigma_{true}$, the true covariance of a given design matrix $X$. 
$\Sigma = C_{ZZ}+\left(|\Lambda_{min}| + \lambda(1-\alpha)\right)I$ is
the approach suggested 
in this paper and $\Sigma = I$ and $\Sigma  = \Sigma_{true}$ are
presented for reference. 
The approach using $\Sigma = I$ is equivalent to imputing the missing 
values with the corresponding column mean of a training set, which ignores
 the correlation structure of the design matrix. The missing pattern and the
 correlation structure of the test set are the same as those of its training set in
 each scenario and every test set has 100 data points.
 Both cross validation and test error are evaluated over 50 trials. \\
\indent Table \ref{tab:table2} and figure \ref{fig:figure2} show that in terms of MSE, the combined
method and the non-negative definite covariance approach outperform
the mean imputation in most cases. The combined method is effective
especially when missing rate is uniform over features and the non-negative
definite method surpasses other two methods when missing rate is high
on signals. 
Figure \ref{fig:figure2} shows that in some cases the global minimum was achieved at $\eta \in (0,1)$ which implies the efficacy of the balancing parameter $\eta$.\\
\indent Table \ref{tab:table3} and table \ref{tab:table4} show that in the combined method with the design matrix having correlation,
imputing the incomplete test set using conditional expectation with
estimated $\Sigma_{est}$ is effective. 
For evaluating test error, the suggested method($\Sigma_{est}$) yields smaller test errors than just imputing missing values with corresponding column means($\Sigma=I$) of the training set.
Also, for choosing the optimal parameters, $\Sigma_{est}$ shows consistently better results.
\section{Conclusion}

This paper discusses the problem of applying penalized regression
when observations are absent.
We first proposed the {\em non-negative definite covariance} approach, which
 forms an unbiased estimator of the objective function
 and then modifies it to ensure convexity.
We extended this approach by combining with the mean imputation method.
We also discussed practical issues such as choosing the
 optimization parameters and predicting $\hat{y}$ in case test
 observations are incomplete.\\
\indent Further investigation of these estimators and their
properties would be valuable, espectially in big data settings.

\newpage
\bibliographystyle{agsm}

\bibliography{draft5_1}
\nocite{*}

\end{document}